\documentclass[11pt]{article}

\usepackage{xcolor,amsmath,natbib}
\usepackage[hmargin=1in, vmargin=1in]{geometry}

\usepackage[colorlinks=true,linkcolor=blue,citecolor=blue]{hyperref}

\begin{document}

\title{Bridging Breiman's Brook:  \\ From Algorithmic Modeling to Statistical Learning}

\author{Lucas Mentch\footnote{University of Pittsburgh, lkm31@pitt.edu} and Giles Hooker\footnote{Cornell University, gjh27@cornell.edu}}

\maketitle

\begin{abstract}
In 2001, Leo Breiman wrote of a divide between ``data modeling'' and ``algorithmic modeling'' cultures. Twenty years later this division feels far more ephemeral, both in terms of assigning individuals to camps, and in terms of intellectual boundaries. We argue that this is largely due to the ``data modelers'' incorporating algorithmic methods into their toolbox, particularly driven by recent developments in the statistical understanding of Breiman's own Random Forest methods. While this can be simplistically described as ``Breiman won'', these same developments also expose the limitations of the prediction-first philosophy that he espoused, making careful statistical analysis all the more important. This paper outlines these exciting recent developments in the random forest literature which, in our view, occurred as a result of a necessary blending of the two ways of thinking Breiman originally described.  We also ask what areas statistics and statisticians might currently overlook.
\end{abstract}


\section{Introduction}
Twenty years after its initial publication, Breiman's ``Two Cultures'' essay \citep{Breiman2001TwoCultures} remains an entertaining and enlightening read, though we are struck by how uncontroversial new readers in the modern era may likely find many of its prescriptions.  We begin our commentary by comparing Breiman's views at the time with current perspectives on data science and argue that while the field of statistics may have fallen short on some fronts, Breiman would no doubt be pleased with a number of strides made in recent decades.  In our view, these developments have been made not because of a switch in statistical thinking from data to algorithmic modeling, but because rigorous statistical analysis has allowed these methods to be understood within more traditional statistical frameworks. We detail how the willingness to expand the statistical toolbox without sacrificing fundamental principles has led to a rapid development of important recent studies on random forests and urge the statistical community to continue these efforts to better understand cutting-edge methodologies in deep learning.  Throughout the remainder of this paper, quotes and references including only a page number are in reference to Breiman's original essay \citep{Breiman2001TwoCultures}.
\section{Data Science:  Breiman's Final Prediction}
``\emph{If our goal as a field is to use data to solve problems, then we need to move away from exclusive dependence on data models and adopt a more diverse set of tools.}"
-- Page 199 \\

The above quote that appeared in the abstract of Breiman's essay provides a reasonably accurate summary of thoughts that he expounds upon in the pages that follow.  To most modern data analysts, however, the aim of ``using data to solve problems'' almost certainly feels like the credo of what we would now refer to as `data science', of which the field of statistics constitutes only a proper subset.  In the decades following the publication of Breiman's essay, the interest in solving problems via data analyses has continued to explode along with the size and scale of the data available.  Predictably, this has led to the current climate in which researchers from across at least dozens of academic fields might rightfully claim to be working on projects intersecting with data science.

In our view, however, the field of statistics has seen substantial growth in its own right.  Machine learning, a field described by Breiman in 2001 as having ``occurred largely outside statistics ... mostly [by] young computer scientists'' (p.\ 200) is now a standard research interest listed on the majority of CVs by young statisticians on the faculty job market. We note that while this interest sometimes includes topics quite far from what would have been described as Machine Learning in 2001, the eagerness on the part of statisticians to plant their flag in the field is a promising development in the eyes of those sympathetic to Breiman's arguments for a more intense focus on algorithmic modeling . Leading statistics journals like The Annals of Statistics, Biometrika, The Journal of the American Statistical Association (JASA), and The Journal of the Royal Statistical Society (JRSS), have shown a willingness, if not an enthusiasm, to publish advances in machine learning research so long as the core contributions are fundamentally grounded in statistics.  At the same time, the Journal of Machine Learning Research (JMLR) has been established that regularly contains individuals on the editorial staff who are unmistakable statisticians, providing researchers a another top-tier journal venue.  Measures of model accuracy as measured via test or cross-validation error -- an aspect Breiman felt was distinctly lacking (p.\ 204) -- are now commonplace in modern statistical publications.


Individuals and departments have also been outspoken in embracing the data science revolution.  Bin Yu, in her 2014 IMS Presidential address, urged statisticians to proudly proclaim themselves data scientists, arguing that while statistics may not \emph{be} data science, no one does more of the job \citep{BinYuIMS}.  Top statistics departments in the United States including, for example, Carnegie Mellon, Yale, and Cornell University, now include `Data Science' in the official department name.  Thus, while the field of statistics has perhaps not expanded quickly enough in the minds of some, we think that Breiman would be very pleased with the willingness to embrace new directions exemplified in recent decades.

\section{Merging Data and Algorithmic Modeling}
``\emph{There is an old saying ``If all a man has is a hammer, then every problem looks like a nail." The trouble for statisticians is that recently some of the problems have stopped looking like nails.}"
-- Page 204 \\

The foundation of Breiman's arguments lie in the distinction between what he refers to as ``data modeling'' as opposed to ``algorithmic modeling''.  Data modeling, according to Breiman, is the traditional statistical modeling context wherein a basic framework like $Y = f(X)+\epsilon$ is assumed and simple parametric models like linear regression, logistic regression, or the Cox model are utilized to estimate $f$.  The algorithmic modeling culture, by contrast, utilizes tools like decision trees and neural networks and ``\emph{considers the inside of the box complex and unknown. Their approach is to find a function $f(x)$ -- an algorithm that operates on $x$ to predict the responses $y$}" (p.\ 199).  ``\emph{The commitment to data modeling}," Breiman argues, ``\emph{has prevented statisticians from entering new scientific and commercial fields where the data being gathered is not suitable for analysis by data models}" (p.\ 200).  ``\emph{The best available solution to a data problem might be a data model; then again it might be an algorithmic model}" (p.\ 204).

But while the distinction between data and algorithmic models may have seemed natural at the time, 20 years later, it feels arbitrary and unnecessary: $Y = f(X) + \epsilon$ \emph{is} a data model!  Throughout the essay, Breiman seems to repeatedly conflate ``data models" and ``parametric models''.  It's not clear why Breiman seemed to feel that tools like decision trees cannot or should not be implemented in conjunction with at least some minor model assumptions, and his reference to Grace Wahba's foundational, and very statistical, work on nonparametric smoothing as outside of mainstream statistics, at least in 2001, seems strange.

Breiman himself goes so far as to admit that one the primary goals in statistics is to obtain information about the underlying data generating mechanism (p.\ 203, 214).  And insofar as this goal is concerned, it's important to stress that neither the data or algorithmic modeling approach is of any value when carried out at the extreme.  A data model, though perhaps interpretable, holds no value if the model itself is not reflective of the process that generated the original data.  On the other hand, an algorithmic model that amounts to an impenetrable black-box, while it may produce accurate predictions, is merely a potentially more efficient version of nature and therefore of limited scientific value -- we haven't learned anything about the mechanisms and variables driving those predictions.

%

In our view, the enormous progress that has been made by statisticians in recent decades came not from abandoning the traditional principals of data modeling, but by merging the cultures Breiman describes.  The willingness on the part of statisticians to expand the traditional modeling toolbox without sacrificing core data analytic and uncertainty quantification principles has led to tremendous gains in our knowledge and understanding of modern supervised learning techniques.  Random forests -- the apple of Breiman's eye -- is certainly one of the methodologies that has benefitted most from this merged mindset.

%

\section{Advancements with Random Forests}
`\emph{So forests are A+ predictors. But their mechanism for producing a prediction is difficult to understand. ... Using complex predictors may be unpleasant, but the soundest path is to go for predictive accuracy first, then try to understand why.}''
-- Page 208 \\

The second sentence in the above quote is perhaps the best summary of the later half of Breiman's essay and arguably the fundamental principle that has stimulated the recent progress in understanding the success and inner workings of black-box learning algorithms like random forests.  At the time, much of the traditional statistical modeling culture revolved around formulating parametric models and making continual tweaks until the model seemed to fit best.  The process begins with an explicit and interpretable model and modifications are made to try and improve accuracy.  Breiman -- stemming from his belief that ``\emph{...the models that best emulate nature in terms of predictive accuracy are also the most complex and inscrutable.}" (p.\ 208) -- urged statisticians to consider the process in reverse:  begin with a an accurate but potentially obscure model and perform follow-up investigation to uncover the impact of individual predictors on the response.

Breiman kickstarts these efforts in his work with random forests and out-of-bag (oob) variable importance, demonstrating that ``\emph{a model does not have to be simple to provide reliable information about the relation between predictor and response variables}" (pp.\ 209-211).  Interestingly, in doing so, he himself implicitly begins to utilize traditional data modeling thinking in order to extract insight from complex models.  In the following sections, we detail how similar thinking has recently led to a rapid advancement in our understanding of random forests as well as the ways in which they can be reliably used to further our understanding of the underlying processes.

%
%
%

\subsection{Statistical Understanding of Random Forests}
``\emph{My kingdom for some good theory.}'' -- Page 205 \\

Breiman's preference for algorithmic models did not lead him to abandon statistical theory; indeed his essay explicitly recognizes the theoretical developments underpinning both smoothing splines and support vector machines (although it unaccountably ignores a large statistical literature on other nonparametric smoothing methods) and he sketched such an analysis of random forests.  The two decades since then have seen statistics slowly fill in theoretical details particularly about the random forest method. It's unclear whether Breiman's essay, or their reliable ``off-the-shelf'' performance (or a combination of the two) that motivated a statistical focus on random forests in particular, but our own interest in these methods was also influenced by recognizing an appealing macroscopic structure that could be used to develop mathematical theory.

The pace of these developments at times felt glacial, due somewhat to the difficulty of developing an analysis that accounts for the greedy splitting strategies involved in building trees. Theoretical headway has come from making a number of modifications, some arguably beneficial, others purely focussed on mathematical tractability:
\begin{enumerate}
\item Methods that provide guarantees on the (geometic) size of the leaves of individual trees, usually obtained by introducing randomly chosen splits. Originally, this involved generating tree structures largely at random \citep{Biau2008,Biau2010,Biau2012}. More recently, the {\em regularity} conditions of \cite{Wager2018} can be enforced by choosing a split at random with some probability while otherwise allowing data-informed splitting methods to be used. Controlling leaf dimensions provides a similar control over within-leaf bias, where the tree represents the underlying relationship as constant and hence forms an important component of consistency results.

\item Methods that enforce independence between tree structures and leaf values. We think of this as avoiding the problem of trees ``chasing order statistics'' or finding structures that isolate exceptionally (and randomly) large values of the response. Choosing tree structures completely at random and letting the data dictate leaf values naturally provides this and facilitates a representation of random forests as kernel methods \citep{Scornet2016}. \cite{Wager2018} propose splitting the data between samples used to determine tree structure and those assigning values to leaves.  The theoretical advantage of this separation is that it allows the values given to leaves to be unbiased for the average response in the leaf. This completes a control of bias when combined with controlling the size of the leaves. The authors also argue for performance improvements in addition to theoretical guarantees based on reducing edge bias. However, this improvement is counteracted by reducing the size of the sample available for building trees (and hence their depth) and we thus regard that trade-off as unclear.  The standard CART split criterion \citep{CART} seems to remain the overwhelmingly most popular choice in practice.

\item Changing the sampling structure. Breiman's original random forest algorithm obtained trees from bootstrap samples of the data. The last five years has seen a switch to employing subsamples instead. This is motivated by the connection between random forests based on subsamples and classical $U$-statistic constructions and allowed both \cite{Mentch2016} and \cite{Wager2018} to develop central limit theorems for the predictions that random forests give. These have been refined by analyses in \cite{Peng2019} who also provided Berry-Esseen results and \cite{Zhou2019} who extended the analysis to subsampling {\em with} replacement and thereby obtained improved variance estimates.  \cite{Scornet2015} also utilize subsampling to obtain what is arguably the ``best" consistency result to date, in that it covers estimators very closely resembling the form of random forests originally proposed by \cite{Breiman2001}.

    The switch to subsampling can be regarded as assisting Breiman's objective of reducing correlation between trees -- we'd like subsamples large enough that trees can be built to reasonable depths (to improve tree-level accuracy) but small enough that between-tree correlation is mitigated.
    While this creates another tuning parameter, \cite{Zhou2019} found that using subsamples with replacement also reduced the sensitivity of results to this fraction, and that the optimal subsample size is not always that of the training data.

    We note that, similar to more classical non-parametric smoothing results, these central limit theorems are not necessarily centered at a target $f(X) = E(Y|X)$. The $U$-statistic argument applies quite generally to averages of any sort of learner, but specific analysis is required to make $E\hat{f}(X) - f(X)$ small enough to be ignorable for inference \citep[e.g.][]{Wager2018}. In nonparametric smoothing, removing the bias from a central limit theorem generally involves under-smoothing relative to the optimal predictive model \citep{eubank1999nonparametric}. While we readily acknowledge the value in being able to perform strictly formal inference (see below for examples) we also believe that an analysis of stability has value in itself.
\end{enumerate}

These developments allow us to provide results that state, for a given point of interest $x$,
\begin{equation} \label{eq:clt}
\zeta_n(x)^{-1}( \hat{f}_n(x) - \eta_n(x) ) \sim N(0,1)
\end{equation}
where particular expressions for the mean and standard deviation functions $\eta_n(x)$ and $\zeta_n(x)$ depend on particulars of the modification of the random forest algorithm. However, the general framework creates the basis for the formalized inference we describe below.

None of the above modifications are strictly necessary in the sense that \eqref{eq:clt} is provably false without them.  As noted above, \cite{Scornet2015} have produced a consistency result for estimators nearly identical to the original random forest algorithm. Such minor modifications have largely been motivated as a means to mathematical tractability, although in some cases they may also provide genuine improvements in predictive or inferential performance.  These ideas have also been accompanied with a wide range of modifications to random forests that repurpose them for estimating individual treatment effects \citep{Wager2018}, survival analysis \citep{gordon1985tree,segal1988regression,davis1989exponential,hothorn2004bagging,zhu2012recursively,steingrimsson2016doubly}, quantile regression \citep{Meinshausen2006} and multivariate outcomes \citep{segal2011multivariate}, with similar analyses to those we outline above.  Breiman's ideas about using co-in-leaf status as a measurement of proximity between points have also been taken up to provide local models \citep{athey2019generalized}, prototypes \citep{tan2020tree} and prediction intervals \citep{lu2019unified}; see also \cite{zhang2019random} for prediction intervals based on conformal methods.

These results have all been focussed around random forest-like methods, but they are hardly the only machine learning methods available.  Though not shy about promoting his own work, Breiman's essay specifically discusses boosting \citep{freund1996experiments,freund1997decision,friedman2001greedy} as a rival methodology with real potential. These methods received early attention from analogies to the LASSO in linear models \citep{friedman2001greedy} and initial analysis in \citet{buhlmann2002consistency}. Boosting is able to accommodate a wide array of model forms and restrictions (Breiman's comments suggest he might not approve of the latter), some of which are explored in \citet{hothorn2010model,lou2013accurate,zhou2018boulevard}. Nonetheless, theoretical analysis remains rudimentary and a long way from \eqref{eq:clt}. Early analyses in this direction in \cite{zhou2018boulevard} try to make boosting look more like random forests and may provide a starting point; Bayesian methods \citep{chipman2010bart,rockova2019ons,hill2011bayesian} have also seen significant development.


\subsection{Real-world Understanding from Random Forests}
``\emph{Framing the question as the choice between accuracy and interpretability is an incorrect interpretation of what the goal of a statistical analysis is. The point of a model is to get useful information about the relation between the response and predictor variables. Interpretability is a way of getting information. But a model does not have to be simple to provide reliable information about the relation between predictor and response variables; neither does it have to be a data model.}'' -- page 209 \\

While statistical understanding of the properties of random forests has focused on analyzing models that are increasingly similar to Breiman's original ideas, this has not been the case for the tools that he advocates for deriving knowledge from them.  He acknowledges that the tools he advocates result in algebraically-complex models that do not allow human inspection, but proposes that we should first find a model that predicts well and then reverse-engineer what it does.  This proposal has been directly attacked by some (e.g.\ \cite{rudin2018please}), but the specifics of Breiman's approaches have  also been shown to be potentially misleading. In fact, identifying, or testing, the important covariates in a random forest is a challenging procedure with many popular methods exhibiting distinct flaws. Here we review some of these and point to ways in which variable importance can be validly assessed, and point to how improved statistical understandings of random forests have contributed to these recommendations.

Breiman's original proposals for eliciting information about the relation between predictor and response variables was via his out-of-bag variable importance measures. These proceed on a tree-by-tree basis:
\begin{enumerate}
\item Choose those data points that were not included in the sample that the tree constructed. Call this data set $Z$.

\item Form a new data set $Z^{\pi}$ by permuting the values of the variable of interest ($X_1$ throughout the discussion below, with $X = (X_1,X_{-1})$).

\item Compare the predictive accuracy of using the current tree to predict the response in $Z$ to that when using $Z^{\pi}$.
\end{enumerate}
The idea of permuting features can be applied generally, but specifically permuting out-of-bag data does, by definition, require an ensemble structure constructed via bootstrapping or at least some form of resampling. It also means we measure accuracy of individual members of the ensemble, not the ensemble as a whole.  This general approach was critiqued by numerous studies including both \cite{Strobl2008} and \cite{Hooker2007} for also breaking any relationships between $X_1$ and $X_{-1}$. In particular it may lead to values in $Z^{\pi}$ that are far from any observed data points, thereby evaluating the ways in which a random forest extrapolates at least as much as the signal that it picks up. Despite these early warnings, the method proved (and unfortunately remains) popular as a variable screening tool; a stereotypical example can be found in \cite{diaz2006gene} where features measured as least important are successively removed.  This prompted us to repeat and formalize the critique with further studies in \cite{Hooker2019}. In those studies, as in many of the earlier works, we found that pairs of correlated features were likely to have their importance estimates inflated, despite having an impact that would conventionally be regarded as smaller. We stand by our recommendation not to use these methods.

The other common measure of variable importance among random forest implementations is specific to tree structures and measures the improvement in accuracy that is obtained each time the tree is spit \citep[e.g.][]{friedman2001greedy}; a variable accumulates the improvements associated with those splits that use it. This has been critiqued for favoring variables with more potential split points in \cite{Strobl2007}\footnote{This critique also motivated the GUIDE procedure \citep{loh2002regression}.} and we admit to having been surprised by the size of the effect found in simulation in \citet{zhou2019unbiased}. This, along with other papers \citep{li2019debiased,loecher2020unbiased} have attempted to correct for this bias by using out-of-sample data, but the precise interpretation of this measure is also not clear outside of the very specific cases in \citet{scornet2020trees}.

Some of the flaws in the above methods are surprising while others, in retrospect, appear foreseable. So are there ways to more appropriately and rigorously assess variable importance? Yes, though none are as computationally or algorithmically simple as those above.

The primary problem with the out-of-bag approaches outlined above is that models are constructed only on the original data -- inspections are made by altering (permuting) test data, but never by altering the training data and reconstructing the model.  We refer to this general idea as ``permute-and-repredict" in \cite{Hooker2019}.  In most practical applications, however, when one asks whether a particular feature -- say $X_1$ -- is ``important", they are really asking whether the model estimate (learning procedure, algorithm) could be just as accurate without it.  Investigating this question directly thus necessarily involves the more computationally intensive idea of rebuilding the model under alterations to the training data.

In our earlier work on random forests \citep{Mentch2016}, we proposed a hypothesis test for variable importance based on exactly this idea:  we construct one random forest on the original data, then another where $X_1$ is removed from the dataset and compare the difference in predictions between the two models.  Noticing that predictions are sometimes affected by the dimension of the feature space, we advocated that the test be carried by permuting $X_1$ in the data used by the second random forest rather than removing it from the model entirely.  In more recent work, \cite{Coleman2019} proposed a more computationally efficient version of this test that measures the difference in accuracy between the two forests rather than the difference in raw predictions.  This kind of idea as also appeared in the conformal inference literature where \cite{lei2018distribution} proposed \emph{LOCO} (Leave-One-Covariate-Out) tests that follow exactly the same intuition and in a model-agnostic test recently proposed in \cite{williamson2020unified}.

While the ``drop-and-rebuild" and ``permute-and-rebuild" approaches both improve upon the ``permute-and-repredict" framework, neither is perfect in practical settings because in addition to breaking the relationship between the feature and response, we also break the relationship between the feature and all remaining features.  A natural further fix is thus to replace $X_1$ with a copy generated from its distribution {\em conditional} on $X_{-1}$. This was, to our knowledge, first advocated in \cite{Strobl2008} and holds some resemblance to the new literature on knockoffs \citep{barber2015controlling}, conditional randomization tests \citep{candes2018panning,liu2020fast}, and holdout randomization tests \citep{tansey2018holdout}.  However, while such an approach is certainly theoretically appealing, the task of generating conditional simulations of $X_1|X_{-1}$ is non-trivial; a number of potential methods exist to do this, but more work needs to be done here and we anticipate this research area will remain very important in the coming years.

Finally, we note that the definition of variable importance itself remains somewhat murky and, much to our frustration, commonly undefined in practical applications.  Conditional generation implicitly suggests that we are primarily interested in  the predictive ability of $X_1$ {\em beyond that provided by the other covariates}.  We could, instead, simulate $X_{-1}|X_1$, or some subset of the covariates. To this end, Shapley values \citep{Lundberg2017unified} have been suggested as representing the marginal improvement in predictive accuracy given by $X_1$, averaged over the order of inclusion. Alternative functional ANOVA methods \citep{Hooker2007}, relate variable importances to statistically-familiar types of sums of squares, allow the examination of interactions and can account for covariate distributions, albeit at a potentially prohibitive computational cost.

\subsection{Formalized Inference from Random Forests}
``\emph{an algorithmic model can produce more and more reliable
information about the structure of the relationship between inputs and outputs than data models.}'' -- page 200 \\

While statistical theory and insight has helped to understand and improve our methods of extracting information from a random forest, we can go further and start to use random forests explicitly for statistical inference.  The CLT in \eqref{eq:clt} can be used to provide confidence intervals about individual predictions \citep{Mentch2016} or other quantities averaged over some covariate distribution \citep{Wager2018}. However, ideas such as variable importance suggest hypotheses about structure along the lines of
\begin{equation} \label{eq:h0}
H_0: \  \eta(X_1,X_{-1}) = \eta^*(X_{-1})
\end{equation}
for some function $\eta^*$. Here we can use the extension of \eqref{eq:clt} to multivariate or process distributions over the input space. In particular, \cite{Mentch2016} provide formal tests of this form by evaluating the differences
\begin{equation} \label{eq:diff}
D(X) = \frac{1}{B} \sum (T_b^{\omega}(X) - T_b^{\pi}(X))
\end{equation}
between trees $T^{\omega}$ build with the original data, and $T^{\pi}$ that used a permuted version of $X_1$. Since the $D(X)$ have a joint multivariate central limit theorem, we can form a $\chi$-square test of their values at a set of equation points.

Alternatively, the functional ANOVA methods in \cite{Hooker2007} were turned into formal tests in \cite{Mentch2017}. Specifically, we interpret the hypothesis in \eqref{eq:h0} as
\[
H_0: \ \eta(X_{i,1},X_{k,-1}) = \eta(X_{j,1},X_{k,-1})
\]
over a set of covariate values $(X_{i,1},X_{i,-1})$. This creates a grid of function evaluations for which $H_0$ is a linear contrast that can be again tested via a multivariate $\chi$-square statistic.  The same techniques can be extended to examine interactions by testing hypotheses of the form
\[
H_0': \ \eta(X_{i,1},X_{j,2},X_{k,-(1,2)}) = \eta_1(X_{i,1},X_{k,-(1,2)}) + \eta_2(X_{j,2},X_{k,-(1,2)}).
\]

Any of these testing frameworks require evaluation at a large number of covariate values, making variance estimates unstable, and \cite{Mentch2017} resorted to random projection methods \citep{srivastava2016raptt} to stabilize them. They also rely heavily on asymptotic normality results and on a fixed set of evaluation points.  \cite{Coleman2019} considered using the two forests that contribute to \eqref{eq:diff}, but developed a permutation test based on randomly swapping trees between forests and measuring the change in differences in error, meaning that far fewer trees need to be constructed, thus allowing for very large test sets.

We hope to show by these methodological developments that Breiman's notions of ``algorithmic modeling'' can be accommodated and used readily within classical statistical paradigms, and refer to \cite{chipman2010bart} for parallel Bayesian developments. Some of the methods have described lean heavily on at least the ensemble structure of random forests, but others are more general provided a result of the form of \eqref{eq:clt} is available. When ``algorithmic models'' are understood as non-parametric regression methods, the chasm that Breiman perceived within statistics can be crossed with only a few well-placed stepping stones. 

\subsection{Why do Random Forests Work?}
``\emph{But when a model is fit to data to draw quantitative conclusions... the conclusions are about the model’s mechanism, and not about nature’s mechanism.}'' -- Page 202. \\

Despite the enormous progress made on the statistical front in recent years, the question that has perhaps proved the most elusive is also the simplest:  Why do random forests seem to work so well in practice?  In their recent review paper, \cite{Biau2016} note that ``present results are insufficient to explain in full generality the remarkable behavior of random forests."  And indeed, studies have long demonstrated the surprisingly robust accuracy of the random forest method.  The most recent and impressive large-scale empirical comparison of methods was provided in \cite{Fernandez2014} where the authors compare a total of 179 classifiers across 121 datasets -- the entirety of the UCI database \citep{uci} at the time.  They found that not only did random forests perform the best overall, but 3 of the top 5 classifiers were some variant of random forests, leading them to declare that ``the random forest is clearly the best family of classifiers."

Until very recently however, few if any explanations for random forest success had been offered that pushed beyond Breiman's original intuition of reducing the between-tree correlation.  \cite{Wyner2017} focused on classification settings and hypothesized that their success was due to interpolation -- a kind of purposeful overfitting that, in their view, was beneficial because perhaps real-world data are not as noisy as statisticians generally believe.

\cite{Mentch2020} were critical of this explanation, noting in particular that in regression settings, the opposite appears to be true:  random forests appear to be \emph{most} useful in settings where the signal-to-noise ratio (SNR) is low.  The authors proposed an alternative explanation based on degrees of freedom wherein the additional node-level randomness in random forests implicitly regularizes the procedure, making it particularly attractive in low SNR settings.  Based on this idea, they demonstrate empirically that the advantage of random forests over bagging is most pronounced at low SNRs and disappears at high SNRs; when the trees in random forests are replaced with linear models, these kind of implicit regularization claims and can be proved formally.  This finding deals a serious blow to the widely held belief shared by both \cite{Breiman2001} and \cite{Wyner2017} that random forests simply ``are better" than bagging as a general rule.  Perhaps most surprisingly, the authors also demonstrate that when a linear model selection procedure like forward selection incorporates a similar kind of random feature availability at each step, the resulting models can be substantially more accurate and even routinely outperform classical regularization methods like the lasso.

 In follow-up work, \cite{Mentch2020AugBagg} build on this implicit regularization to introduce another surprising result:  the predictive accuracy of both bagging and random forests can sometimes be dramatically and systematically improved (particularly at low SNRs) by including additional noise features in the model which, by construction, hold no additional information about the response beyond that contained in the original features.  The authors refer to these kinds of procedures as \emph{augmented bagging} and show that even the formal hypothesis testing procedures above will, correctly but counterintuitively, recognize large groups of noise features as making a significant contribution to the predictive accuracy of the model.

 This last point is quite surprising and easy to misinterpret; readers seeing this for the first time might reasonably ask whether these kinds of hypothesis tests should really be trusted if they routinely identify noise features as predictively significant.  We stress, however, that these tests are merely evaluating whether there is a significant improvement in accuracy when additional noise features are added to the model.  This does indeed appear to be the case in some settings -- models with additional noise features really \emph{can} make the models more accurate.  Thus, in those settings, by identifying such noise features as predictively helpful, it's important to realize that these tests are not making an error.  Rather, as discussed at length in \cite{Mentch2020AugBagg}, it is not the tests themselves that are wrong but the common (mis)interpretations of those tests.

 However, this observation does confound two separate questions: whether the feature improved the predictive accuracy of a model, and whether it is associated with the response by real-world processes. Alternatively: is the feature {\em uniquely} helpful: do we need to include the same measurements, or would the current value of the microwave background radiation be just as useful?   "What is your model for the data?" -- the question Breiman objected to -- in fact becomes essential: are we measuring a property of the world that generated the data, or the algorithm used to analyze it? Both are reasonable to measure, but they are not the same, and can't be assessed by the same methods. Crucially, Breiman's standard of predictive accuracy, while applying to properties of the algorithm, is not necessarily indicative of real-world processes.
 We expect that deeper studies in the general area of variable importance will continue to remain an important topic of study in future years as data becomes larger and more complex and black-box models are increasingly relied upon as a result.


%
%
%

\section{Illustration}

``\emph{It maybe revealing to understand how I became a member of the small second culture... My experiences as a consultant formed my views about algorithmic modeling.}'' -- page 200 \\

Brieman illustrates the development of his views on how to approach data through a series of examples and David Cox, his strongest critic, also notes the paucity of data that actually appears in the highest profile statistics journals. The conversation between statistical theory and applied practice is of vital importance, particularly in discussing broader questions about modeling philosophy, and we here provide a pr\'{e}cis of a project to which we contributed and which we believe demonstrates well the fusion of both machine learning methods and statistical thinking. The interested reader is referred to \citet{Coleman2020} for full details.

The ebird project at the Cornell Laboratory of Ornithology (www.ebird.org) collects data from amateur birdwatchers across the United States. This data repository has been used to estimate over-all changes in bird abundance, provide maps of bird migration, and explore the dependence of bird distributions on ecological features.  It has been a significant motivation for our own research.  The project we describe used these data to assess the impact of local weather conditions on the timing tree swallows' ({\em Tachycineta bicolor}) fall migration. Specifically, daily maximum temperature was expected to drive tree swallow migration through its (unobserved) effect on the abundance of flying insects that make up the greatest part of their food source.

Modeling migration effects is made challenging by complex spatio-temporal effects along with high-order interactions among other landscape and land-use features \citep{zuckerberg2016novel}. This makes machine learning models particularly useful, and these have be extensive explored, for example in \cite{fink2020modeling}. In this project we had a data set of 25727 observations collected after day 200 on each year from 2008 through 2013 in six wildlife refuges along the north eastern United States coast. These observations recorded the presence or absence of tree swallow sightings, the time and day of the recording, and the time and distance effort put into data collection. We also collected local landcover and elevation summaries from remote sensing systems as well as the daily maximum temperature, providing a total of 30 variables for each observation.

Within this data set, daily maximum temperature is strongly associated with day of year to the extent that there is little evidence of its predictive utility. This relationship also makes permuting the maximum temperature values problematic in creating unrealistically warm December days and very chilly Augusts. Instead, we replace maximum temperature with its difference from the average for that day and location over the six year period. This orthogonalization isolates the effect of weather from other temporal drivers, both allowing a machine learning algorithm to make use of the signal and reducing the effect of permuting the difference.

We assessed whether maximum temperature had a discernible affect on tree swallow abundance by examining the differences in predictions between tree ensembles obtained from the original data and those obtained using a permuted version of maximum temperature difference. This was done at 25 time-locations pairs in each of the wildlife refuges in which we used a multivariate version of \eqref{eq:clt} to develop a $\chi^2$ test for each 25-vector of prediction differences. This revealed strong evidence of an effect at 4 of the regions with much attenuated evidence in the most northerly and southerly regions where we expect migration timing and climate to already reduce the role of temperature.

This study illustrates the combining of the modeling cultures that Breiman describes: statistical considerations play a key role in designing the covariates we used, the development of our tests, and our assessment of evidence, but we avoided the need for extensive and complex models by using algorithmic tools: treating these as statistical models is not a large leap.  Debates around causality and fairness in machine learning have equally demonstrated the need to combine schools in this way and we expect the gap that Breiman perceived to become increasingly invisible.

\section{What is the New Algorithmic Modeling?}

``\emph{This is a fascinating enterprise, and I doubt if data models are applicable. Yet I would enter this in my ledger as a statistical problem.}'' -- page 214 \\

Breiman's essay clearly displays his frustration with the conservativism of a statistics community unwilling to engage with unfamiliar ideas and methods. This same sentiment is not unfamiliar to us but it is also hardly unique to statistics, as many statisticians who have received reviews from the computer science conferences promoted by Breiman can attest. We do not subscribe to the sentiment that ``Statistics lost Machine Learning'', but the statement does lead us to ask what other areas we might currently be going overlooked.

An obvious answer is ``deep learning'' (viewed at the time of Breiman's essay as being somewhat played out) and indeed we are aware of starkly little statistical involvement in this intensely hot area of machine learning research  where the greatest performance gains over the past decade -- Breiman's primary metric of validity -- have come from.  The work that we are aware of \citep[e.g.][]{barron2018approximation,bai2020efficient} focusses on shallow networks and largely ignores the algorithms used to learn weights. This seems analogous to the original work on random forests and we hope that it will similarly lead to steadily more practical results, but also expect progress will require new ways of representing the problem.

However, rather than the mathematical properties of neural networks, the aspect we most worry about statisticians ignoring is the type of data where these tools excel: images, text, sound and other natural data types that are both highly structured -- e.g.\ pixels are next to other pixels -- and where individual numbers have little to no individual interpretive value.  (The large statistical literature on medical imagining requires making images comparable on a value-by-value or at least region-by-region level. Similar comments can be made about inference with many other non-Euclidean objects where defining a distance appears to be a starting point.)  We have seen little statistical interest in the way deep learning is applied to these types of complex data objects where questions about manifold structure, the stability of explanation methods, or structuring models that extract information could all benefit from statistical perspectives, even as we freely acknowledge that we, ourselves, have not contributed.

Outside of these specifics, we expect that Breiman would have welcomed the broadening of interests associated with the field of Data Science. Statistics has a long history of worrying about the design or causal issues in data provenance and the biases associated with data collection, and a somewhat more limited engagement with data privacy. Much more of this is sorely needed as the emerging issues in ethical AI are making clear. We think the adjacent areas of data merging and data cleaning are both ripe for more formal study; developing methods and analyzing their consequences. There are surely other tasks of which we haven't thought, but we really do need to encourage the discipline to expand beyond our collective intellectual comfort zone.

\section{Discussion}

``\emph{The roots of statistics, as in science, lie in working with data and checking theory against data. I
hope in this century our field will return to its roots.
There are signs that this hope is not illusory. Over
the last ten years, there has been a noticeable move
toward statistical work on real world problems and
reaching out by statisticians toward collaborative
work with other disciplines. I believe this trend will
continue and, in fact, \textbf{has} to continue if we are to
survive as an energetic and creative field.}"
-- Page 214 \\

Looking around today at the emerging field of data science and the resulting shifts within the statistics community, it's easy to see the impact of Breiman's thinking.  As faculty members in statistics departments who are active in data science education at both the undergraduate and graduate levels, we frequently interact with students who are new to the field, many of whom are reading Breiman's seminal essay for the first time.  In the vast majority of cases, we find that these young statistics students are puzzled as to why it was ever considered a controversial work.  ``\emph{Isn't it obvious that this is what we're supposed to be doing?"} Ironically, perhaps the best indication of Breiman's lasting impact through his ``Two Cultures" essay is its lessened relevance today.

Just as the field of statistics has begrudgingly opened it's mind to problems outside the classical modeling framework, computer scientists and machine learning researchers deserve credit for expanding their horizons as well.  A newfound interest in interpretability (e.g.\ ``explainable AI"), causality, and fairness is commonly on display at each of the most notable machine learning conferences, each of which hearkens back to fundamental ideas and philosophical considerations in traditional statistical modeling.

It's worth noting that while the boundaries of these fields may have softened, it remains the case that their respective directions are not unified.  Computer scientists remain largely driven by the ``what" (predictive accuracy); statisticians the ``how" and ``why."  We still observe a separation of concerns: global interpretations versus local explanations mirrors scientific knowledge versus actions.  In our view, however, so long as we resist dogmatic aversions to alternative ways of thinking, unity shouldn't be seen as a necessity for (or perhaps even the optimal route to) continued progress.

Finally, there are likely some statisticians who, like David Cox's commentary in the original paper, remain skeptical of the recent trends in our field towards data science, fearing that it signifies a surrender of the theoretical school of statistics and traditional modeling.  We do not think so; rather it is the realization that both sides might have something to tell the other.  Rather than seeing these trends as \emph{shifting away from} classical statistical ideas, we urge skeptical colleagues to consider these recent developments an \emph{expansion of} those ideas.  This, we firmly believe, is the central theme of Breiman's essay and is ultimately what allowed for the exciting recent developments detailed in the previous sections.

%

\bibliographystyle{chicago}
\bibliography{database}

\end{document}